\documentclass[]{article}
\usepackage[top=1in, bottom=1in, left=1.25in, right=1.25in]{geometry}
\usepackage{graphicx}
\usepackage{authblk}
\usepackage{chngcntr}

\title{Optimizing the linearity in high-speed photodiodes}

\author[1,*]{J. Davila-Rodriguez}
\author[2]{X. Xie}
\author[2]{J. Zang}
\author[3]{C. J. Long}
\author[1]{T. M. Fortier}
\author[1,4]{\mbox{H. Leopardi}}
\author[1]{T. Nakamura}
\author[2]{J. C. Campbell}
\author[1,4]{S.A. Diddams}
\author[1,$\dagger$]{F. Quinlan}

\affil[1]{NIST, Time and Frequency Division, 325 Broadway MS 847, Boulder, CO 80305, USA}
\affil[2]{Department of Electrical and Computer Engineering, University of Virginia, Charlottesville, VA, USA}
\affil[3]{NIST, Communications Technology Laboratory, 325 Broadway, Boulder, CO 80305, USA}
\affil[4]{ Department of Physics, University of Colorado Boulder, 440 UCB Boulder, Colorado, 80309, USA}

\affil[*]{Corresponding author: josuedavila@gmail.com}
\affil[$\dagger$]{franklyn.quinlan@nist.gov}

\begin{document}

\maketitle

\begin{abstract}
Analog photonic links require high-fidelity, high-speed optical-to-electrical conversion for applications such as radio-over-fiber, synchronization at kilometer-scale facilities, and low-noise electronic signal generation. Photodetector nonlinearity is a particularly vexing problem, causing signal distortion and excess noise, especially in systems utilizing ultrashort optical pulses. Here we show that photodetectors designed for high power handling and high linearity can perform optical-to-electrical conversion of ultrashort optical pulses with unprecedented linearity over a large photocurrent range. We also show that the broadband, complex impedance of the circuit following the photodiode modifies the linearity significantly. By externally manipulating the circuit impedance, we extend the detector's linear range to higher photocurrents, with over 50~dB rejection of amplitude-to-phase conversion for photocurrents up to 40~mA. This represents a 1000-fold improvement over state-of-the-art photodiodes and significantly extends the attainable microwave power by a factor of four. As such, we eliminate the long-standing requirement in ultrashort pulse detection of precise tuning of the photodiode's operating parameters (average photocurrent, bias voltage or temperature) to coincide with a nonlinearity minimum. These results should also apply more generally to reduce nonlinear distortion in a range of other microwave photonics applications.
\end{abstract}

\section{Introduction}
High-speed photodiodes are key components of a wide range of photonic systems such as microwave-photonic links \cite{Urick2011,Cliche2006}, opto-electronic oscillators (OEOs) \cite{Yao1996,Eliyahu2008}, photonic-based radar \cite{Ghelfi2014}, photonic signal processing \cite{Torres-Company2014,Valley2007}, high speed waveform generation \cite{Li2014}, and the generation of low-noise microwaves via optical frequency division (OFD) \cite{Fortier2011,Xie2017}. These systems take advantage of the low loss, low noise, and large bandwidth inherent in optical systems to provide improved performance over their purely electronic counterparts. As the optical-to-electrical converter, high fidelity operation of the high-speed photodiode is critical in limiting excess noise and signal distortion that would otherwise largely undermine the advantages of employing photonic techniques in microwave signal generation, dissemination, and processing.

In many cases, the noise and linearity of a photodetector cannot be considered separately. For example, the output microwave power of a photodetector will saturate with high optical power illumination, limiting the detector's operating range. This can in turn constrain the achievable signal-to-noise ratio (SNR) of a link, and can limit the photonic gain of microwave photonics systems. With the development of high saturation power photodiodes capable of generating microwave power approaching 2~W \cite{Xie2014}, increased SNR and all photonic gain \cite{Devgan2009} have become more feasible to a wider range of microwave photonic systems. Prior to the onset of saturation, nonlinearity is present in the form of frequency mixing, distorting the microwave output with the generation of harmonic and intermodulation distortion (IMD) products. While this can be used advantageously for frequency up- and down-conversion \cite{Cabon2010,Jaro2003,Fushimi2004}, nonlinear mixing generally reduces the spur free dynamic range of photonic links, and, in the case of ultrashort optical pulse detection, result in amplitude-to-phase conversion (\mbox{AM-to-PM}). The \mbox{AM-to-PM} nonlinearity has drawn considerable interest in low noise microwave generation by OFD, where the mode-locked laser's relative intensity noise (RIN) is converted to phase noise \cite{Bouchand2017,Taylor2011,Ivanov2005,Hu2017}, degrading the timing jitter of the microwave signal of interest. As the RIN of a train of ultrashort optical pulses can exceed the pulse-to-pulse timing jitter by orders of magnitude and \mbox{AM-to-PM} conversion coefficients can approach unity, \mbox{AM-to-PM} conversion can easily dominate all other noise sources in the generated microwave signal. As discussed in more detail below, \mbox{AM-to-PM} may be considered a special case of IMD where all of the microwave signals are harmonically related. Thus, studies of and solutions to the problem of \mbox{AM-to-PM} conversion apply to systems beyond those involving ultrashort pulse detection. 

Existing solutions to the \mbox{AM-to-PM} problem in high-speed photodetectors can be separated into two general categories. One is to avoid direct detection of the ultrashort pulses, and instead synchronize a microwave oscillator to the optical pulse train \cite{Kim2006,Lessing2013,Peng2014}. This can be accomplished by way of a Sagnac interferometer containing an electro-optic modulator that compares the optical pulse time of arrival to the zero-crossings of the microwave oscillator. While this technique has demonstrated AM rejection as large as 60~dB \cite{Lessing2013}, the output frequency is limited to the microwave oscillator's capture range, and it hasn't demonstrated the extremely low phase noise capabilities of direct optical pulse detection \cite{Quinlan2013}. The other solution is to directly detect the optical pulses, but operate at an \mbox{AM-to-PM} ``null" point where the coefficient is near zero \cite{Xie2017,Taylor2011,Baynes2015}. By careful adjustments to the illuminating optical power and the photodetector bias voltage, 50~dB rejection of the amplitude onto the phase can be achieved. Here it is the extremely fine tuning, sometimes necessitating active feedback mechanisms \cite{Bouchand2017}, that makes relying on a null point operationally challenging. In this work, we show that charge-compensated modified uni-traveling carrier (\mbox{CC-MUTC}) photodiodes under short pulse illumination support 10~GHz generation with \mbox{AM-to-PM} coefficients below -50~dB over a photocurrent range of 40~mA, an improvement of several orders of magnitude over standard and advanced p-i-n photodiode designs \cite{Bouchand2017,Taylor2011}. This level of \mbox{AM-to-PM} rejection enables the utilization of these photodiodes without additional stabilization schemes and at higher microwave output powers than previously reported. Achieving these results required manipulating the nonlinear interplay between the photodetector and the external microwave circuit. By modifying the broadband complex impedance of the circuit following the photodiode, the system's \mbox{AM-to-PM} is drastically altered. Using such a scheme, we are able to generate 15~dBm (30~mW) of microwave power at 10~GHz with 53~dB rejection of amplitude noise onto the microwave phase. Importantly, this scheme can be readily adopted by other systems using high linearity photodiodes, including those that do not utilize ultrashort optical pulses

\section{Photodiode nonlinearity and cancellation}
In contrast with p-i-n photodiodes, MUTC photodiodes have demonstrated large saturation powers in part by separating the absorber into depleted and undepleted sections \cite{Li2004,Li2010,Nagatsuma2009}. \mbox{CC-MUTC} detectors further improve power handling by pre-distorting the electric field in the depletion region, resulting in microwave power generation exceeding 1~W under modulated-CW light illumination \cite{Xie2014,Li2011,Beling2016,Zhou2013}. This design is also beneficial to short pulse detection, where large peak photocurrents can shape the electric field and alter the detector's response. In p-i-n photodiodes, the induced electric field due to the photocarriers tends to broaden the photocurrent pulse as the average optical power is increased \cite{Diddams2009,Zhang2012}. However, as a consequence of the field pre-distortion and the photocarrier-induced fields in the undepleted regions in \mbox{CC-MUTCs}, the width of the current pulse slightly decreases at moderate average photocurrents, prior to the onset of saturation (where the current pulse width begins to significantly increase). This is depicted in Fig. 1(a). The changing impulse response impacts the \mbox{AM-to-PM} nonlinearity because the pulse ``center of mass" is photocurrent-dependent, leading to a coupling between amplitude and timing. Usually, we are concerned with how this impacts the phase of the microwave signal of interest, represented by a single harmonic of the pulse train repetition frequency. Since each harmonic individually carries the information of the optical pulse train timing \cite{vonderlinde1986}, measuring the \mbox{AM-to-PM} (and the timing jitter) on a single harmonic obviates the more difficult task of handling the full photocurrent spectrum. As illustrated in Figs. 1(a) and 1(b), the changing duration of the current pulse leads to the phase of a harmonic initially decreasing before increasing as the impulse width begins to broaden. The change in delay for a fractional change in the photocurrent is the \mbox{AM-to-PM} coefficient, illustrated in Fig. 1(d). Important parameters in the \mbox{AM-to-PM} curve are the coefficient's value at its peak, the average photocurrent at which the \mbox{AM-to-PM} crosses zero (the null point), and its slope at the zero crossing. These values determine how close to the minimum the photodiode should be operated for a given application, whether the \mbox{AM-to-PM} null is experimentally accessible, and the amount of microwave power that can be achieved with low \mbox{AM-to-PM} conversion. Evidently, the desired condition is a low coefficient value for the largest possible photocurrent and – when the intention is to operate at the null – a smaller slope at the zero crossing such that small changes in the photocurrent don't spoil the \mbox{AM-to-PM} coefficient.

\begin{figure}[htbp]
\centering
\includegraphics[width=4.5 in]{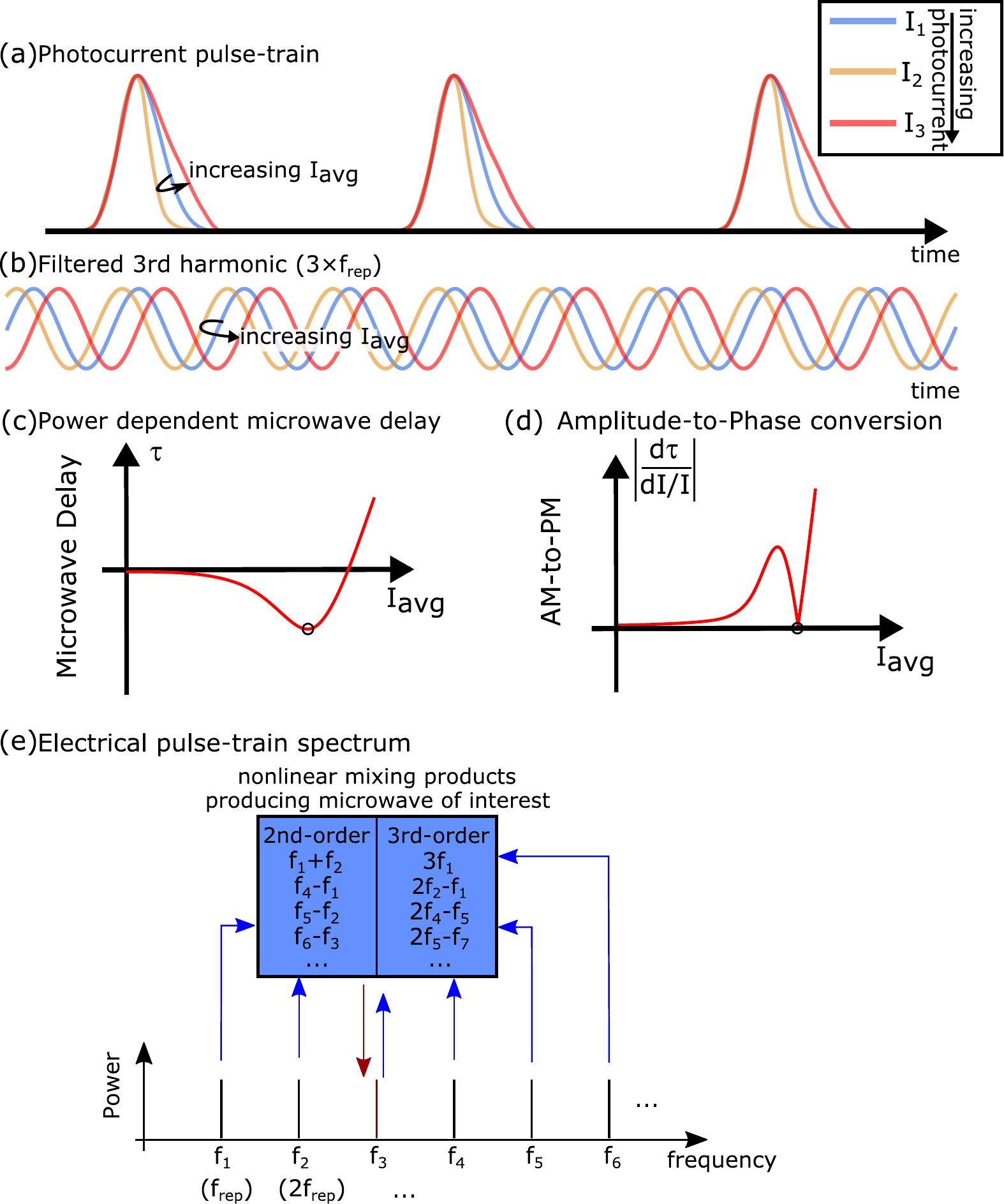}
\caption{The inherent nonlinearity of the photodiode. The carrier dynamics are affected by the electric fields in the device, which change as a function of average photocurrent due to the photocarrier-generated fields. This leads to (a) average photocurrent dependent impulse response which translates into (b) photocurrent dependent microwave phase. (c) Shows the microwave delay as a function of photocurrent, converting (d) amplitude modulation to phase modulation. The width of the impulse response of the photodiode is depicted in (a) to have a minimum at a certain photocurrent, causing the existence of a turn-around point in the phase of the extracted microwave signal and a zero crossing of the \mbox{AM-to-PM} coefficient. This is emphasized by the circles in (c) and (d). (e) Frequency domain picture of the nonlinearity in a diode illuminated by a periodic optical pulse-train. All the tones generated are harmonically related, leading to a myriad of mixing products feeding into the same harmonic of the repetition frequency. The combined nonlinear phase-shifts to the harmonic of interest produce the \mbox{AM-to-PM} conversion shown in (d).}
\label{fig:false-color}
\end{figure}

Physics-based models of photocarrier transport of \mbox{CC-MUTCs} underpin the description above \cite{Hu2017,Xie2017a}. Of course, the particular carrier dynamics of different photodiode structures will vary. However, all photodiodes (MUTC and p-i-n) have displayed some photocurrent value where there is a sign change and a zero crossing in the \mbox{AM-to-PM} conversion value \cite{Taylor2011,Xie2017a}. Typically, the magnitude of the \mbox{AM-to-PM} coefficient varies by 30 to 40~dB as the average photocurrent is changed \cite{Taylor2011,Zhang2012}. The large \mbox{AM-to-PM} coefficient away from the null has led to feedback schemes \cite{Bouchand2017} where the optical power is kept at a constant level to maintain the photodiode's operation close to the null, as well as combining the outputs of two photodetectors with opposite sign nonlinearities \cite{Zhang2014}. As shown here, \mbox{CC-MUTC} detectors with optimized doping concentrations and layer thicknesses can reduce the variations in the carrier velocities, providing exceptionally low \mbox{AM-to-PM} over an extended photocurrent range.

For any photonic application, the photodetector is only one piece that will impact the overall system nonlinearity \cite{Urick2011}. We therefore also explore the importance of the microwave circuit following the detector. To this end, a frequency domain description, as shown in Fig. 1(e), is useful. Nonlinear distortion within the photodiode results in the generation of new frequencies as well as the mixing of microwave frequencies (collectively referred to as intermodulation distortion \cite{Lui1990,Pedro2003,Maas}). In the detection of ultrashort optical pulses, all the generated microwave signals are harmonically related, and \mbox{AM-to-PM} conversion may be viewed as the result of nonlinear mixing of the pulse train harmonics. As the photocurrent is varied, the amplitude and phase of the multitude of mixing products changes, altering the phase (and amplitude) of the microwave harmonic of interest. Additionally, like all diodes, photodiodes respond to changes in their bias voltage according to the exponential voltage-current relation. If microwave power is reflected back into the diode, the bias voltage is modulated thereby mixing the microwave fields directly generated by the optical illumination with the reflected signal \cite{Cabon2010,Jaro2003,Fushimi2004}. The newly generated microwave signal then adds to the original microwave from the optical field, resulting in a microwave phase shift at the output of the circuit. The strength of this effect must depend on the average photocurrent to modify the overall nonlinearity. See supplementary information for more details. 

The significance of the nonlinear phase shift due to a reflected microwave signal depends on the strength of the photodiode's inherent nonlinearity and the size of the reflected signal. Experimentally, we find there is always some reflection of the repetition rate harmonics due to the difficulty in achieving perfect broadband impedance matching. The importance of impedance matching has been noted previously \cite{Bouchand2017,Ivanov2005}, although small unintentional reflections combined with the large native nonlinearity of the photodiodes resulted only in small shifts in the nonlinearity near an \mbox{AM-to-PM} null. However, in photodiodes with exceptionally high linearity such as \mbox{CC-MUTCs}, this effect can be much more prominent and great care should be taken with the complex impedance of the circuit following the photodiode. Here we use the additional nonlinearity to our advantage by purposefully reflecting microwave power back to the diode to modify its behavior, analogous to harmonic load pull optimization in high power transistors \cite{Stancliff1979,Ferrero2013}.

\section{Experiments and Results}
To measure the \mbox{AM-to-PM} coefficient of the photodiodes, we use the pulse-train from a mode-locked Er:fiber laser with a repetition rate of 208.33 MHz. This pulse-train is sent through four stages of fiberized asymmetric Mach-Zehnder interleavers (AMZI) with an arm on each stage having an additional half-period delay for the input pulse-train repetition frequency \cite{Haboucha2011,Jiang2011,Quinlan2014}. After four-stage pulse interleaving, this results in a 3.33~GHz pulse-train which is then sent through a final AMZI with a 100~ps (third-period) asymmetry, producing 3.33~GHz pulse-pairs separated by 100~ps. The final stage boosts the achievable power in the 10~GHz harmonic, which is the microwave frequency of interest. The use of optical pulse interleavers has the additional benefit of reducing the number of harmonics in the photocurrent spectrum to be manipulated, such that the predicted additional nonlinearity is more likely to be observed and the results are easier to interpret. The spectrum of the resulting optical pulse-train is filtered to 1 to 2 THz bandwidth and compressed to nearly Fourier-transform-limited pulse-width of $\sim$1~ps before being photodetected. The pulse-width is measured directly in front of the photodiode using a flip mirror to redirect the light to an autocorrelator. All \mbox{AM-to-PM} measurements presented here are performed on the pulse train harmonic at 10~GHz. 

\begin{figure}[htbp]
\centering
\includegraphics[width=3.3 in]{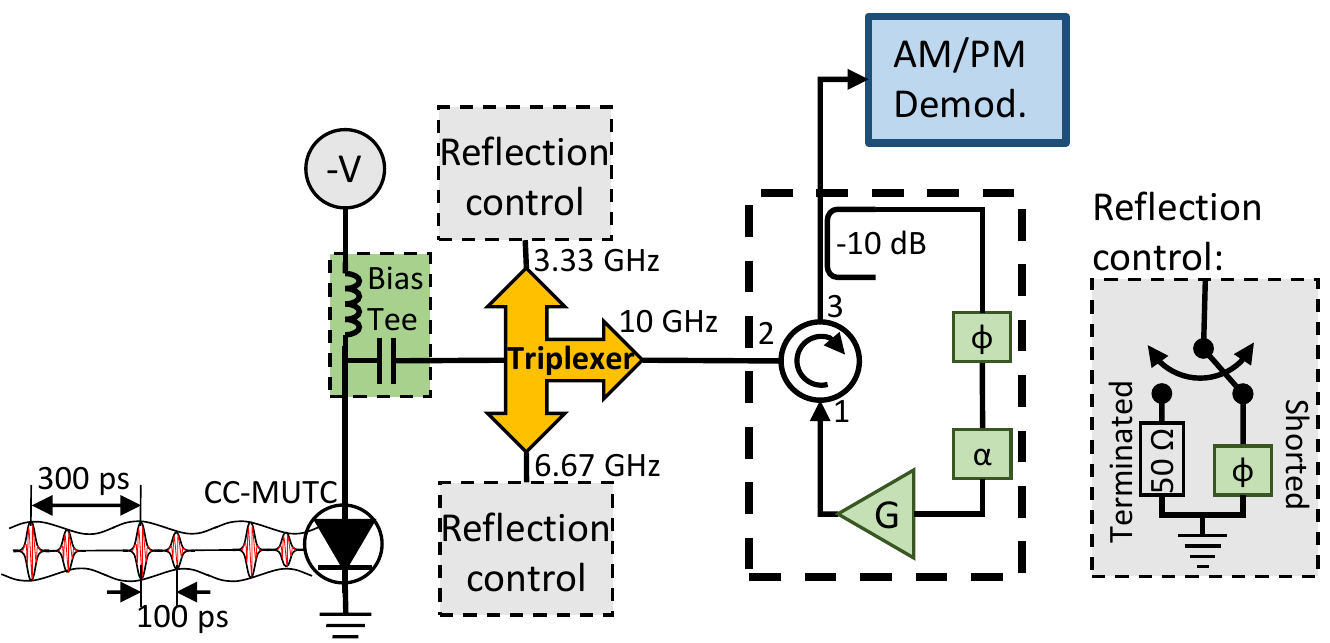}
\caption{Amplitude-to-phase conversion measurement setup. By using a microwave demultiplexer it is possible to have improved control over the reflections of each harmonic present in the circuit. The 3.33~GHz and 6.67~GHz harmonics are reflected using a controllable delay and a short circuit. In the case of the 10~GHz harmonic a fraction of it can also be reflected by using a circulator as shown in the dashed box. The signal is re-amplified in the loop, but the overall gain remains below unity to avoid oscillation. The pulse-train shown in this figure corresponds to the interleaved mode-locked laser pulse-train and is amplitude modulated by an acousto-optic modulator before impinging on the photodiode. Three separate demodulation systems of different type were used in these experiments to verify our results.}
\label{fig:expsetup}
\end{figure}

Just prior to photodetection, the optical pulse-train is amplitude modulated at a low frequency (200 kHz) with an acousto-optic modulator (AOM). The nonlinearity of the photodiode converts some of this amplitude modulation to phase modulation, as described above. To calculate the \mbox{AM-to-PM} coefficient we perform AM and PM demodulation of the 10~GHz microwave and compute the ratio of the measured sidebands. We report the amplitude to single-sideband phase (L(f)) conversion. Explicitly, if the depth of AM is measured in dBam${}^2$ and the PM depth in dBrad${}^2$, then the \mbox{AM-to-PM} coefficient reported here is given by $\textrm{PM} - \textrm{AM} - 3~\textrm{dB}$.

The photodiode's \mbox{AM-to-PM} coefficient was measured as a function of photocurrent both for the case when all the microwave reflections were terminated to the best of our ability, as well as for different chosen values of the reflected phase when one or more microwave signals are intentionally reflected. Observed changes in the \mbox{AM-to-PM} coefficient were consistent with our simple model (further elucidated in the Supplement). We observed similar results across multiple photodiodes and under varied conditions. 

\begin{figure}[htbp]
\centering
\includegraphics[width=3.3 in]{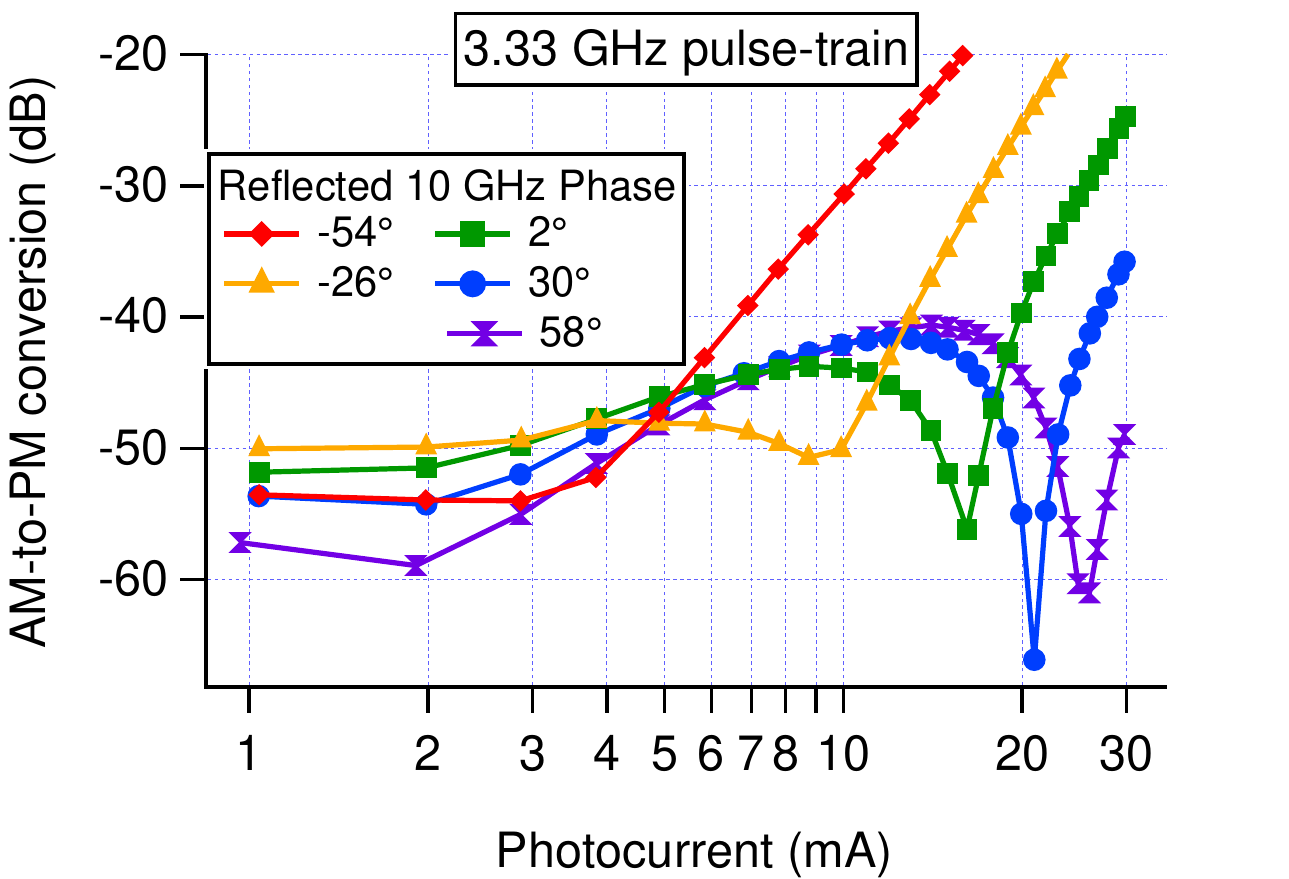}
\caption{Measured \mbox{AM-to-PM} coefficient with the mode-locked pulse-train. In this case the 3.33~GHz and 6.67~GHz harmonics are terminated to 50 $\Omega$ and a small amount of 10~GHz signal is reflected to the diode with a delay via a microwave circulator as shown in the setup in Fig. 2. Notice the shift of the \mbox{AM-to-PM} null to higher photocurrents for phases $>0$ in the reflected signal. Also notice that for phases $<0$ the \mbox{AM-to-PM} coefficient may show a minimum without going through zero or no minimum at all. Each of these cases can be qualitatively explained by our simple single diode mixing model in the Supplement.}
\label{fig:results1}
\end{figure}

To verify the simple model described in the Supplement we reflect a small amount of the 10~GHz harmonic back to the diode via a circulator as shown in the dashed box in Fig. 2, where it can mix with the 20~GHz harmonic and modify the nonlinear phase of the original 10~GHz tone. We then record the \mbox{AM-to-PM} coefficient as a function of photocurrent for several delays of the reflected signal. The introduced relative delays have been carefully calibrated up to an arbitrary overall offset. This overall phase offset has been estimated from a combination of S21 measurements and residual coplanar waveguide lengths. The results are shown in Fig. 3. For clarity, we chose a few representative curves which illustrate the kinds of effects that we expect to observe. For example, for a certain range of phases of the reflected signal, the photocurrent null is expected to appear at progressively higher photocurrents. In Fig. 3 this is represented in the 2$^\circ$ (squares), 30$^\circ$ (circles) and 58$^\circ$ (hourglasses) respectively. On the other hand, when the additional nonlinearity has the same sign as the photodiode's native nonlinearity, they add constructively, and the photodiode performance deteriorates. This is clearly seen in the case of the -54$^\circ$ data shown with diamonds in Fig. 3. Finally, there is the possibility that the additional nonlinearity improves the performance over a small photocurrent range, but it does not allow the existence of a zero crossing. The triangles in Fig. 3 illustrate this case.

\begin{figure}[htbp]
\centering
\includegraphics[width=3.3 in]{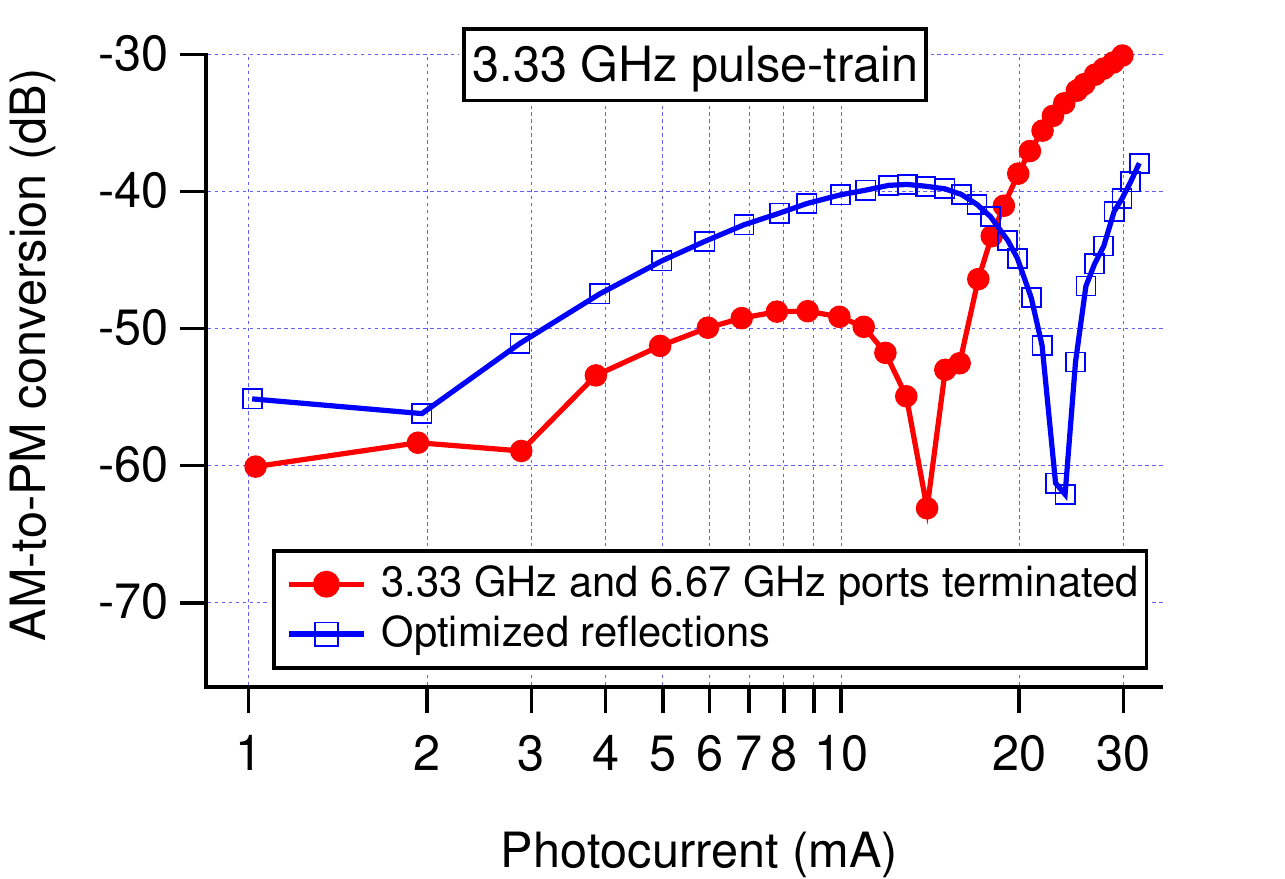}
\caption{\mbox{AM-to-PM} coefficient obtained by ``best effort" termination and by optimizing the reflection phase of the 3.33~GHz and 6.67~GHz harmonics to obtain the lowest AM-to-PM coefficient at 24 mA.}
\label{fig:results2}
\end{figure}

More importantly, we are interested in optimizing the reflection phase to obtain a minimum in the \mbox{AM-to-PM} coefficient at higher photocurrent than under the case with minimized microwave reflections. To show the power of the technique, we set the bias voltage at -12~V and choose 25~mA average photocurrent for optimization. We then proceeded to optimize the reflected phase of both the 6.67~GHz and 3.33~GHz harmonics to minimize the \mbox{AM-to-PM} coefficient. The resulting \mbox{AM-to-PM} curve is shown in Fig. 4. The red circles show the case when we terminate each harmonic and the blue squares show the case when we control the reflection of the 6.67~GHz and 3.33~GHz harmonics. Important features to note are: 1) in the case with terminated harmonics the \mbox{AM-to-PM} coefficient remains below -48~dB up to $\sim$16~mA of photocurrent – already an unprecedented result which showcases the high linearity of the \mbox{CC-MUTC}, 2) relative to the terminated case the \mbox{AM-to-PM} coefficient at 24~mA is improved by $\sim$25~dB due to the additional nonlinearity induced by the microwave reflections, and 3) the photocurrent range with \mbox{AM-to-PM} below -40~dB is extended from 20~mA in the terminated case to 30~mA when the reflection phase of the 3.33~GHz and 6.67~GHz harmonics is optimized. As mentioned above, the effect of reflected microwaves modifying the \mbox{AM-to-PM} coefficients has been observed before, but typically it consisted of a small shift in the location of the null. We observe more significant changes in the coefficient because the reflections have an effect when the inherent \mbox{AM-to-PM} coefficient of the photodetector is sufficiently close to zero. In \mbox{CC-MUTC} diodes, there is a wide photocurrent range where this condition is met. 

\begin{figure}[htbp]
\centering
\includegraphics[width=3.3 in]{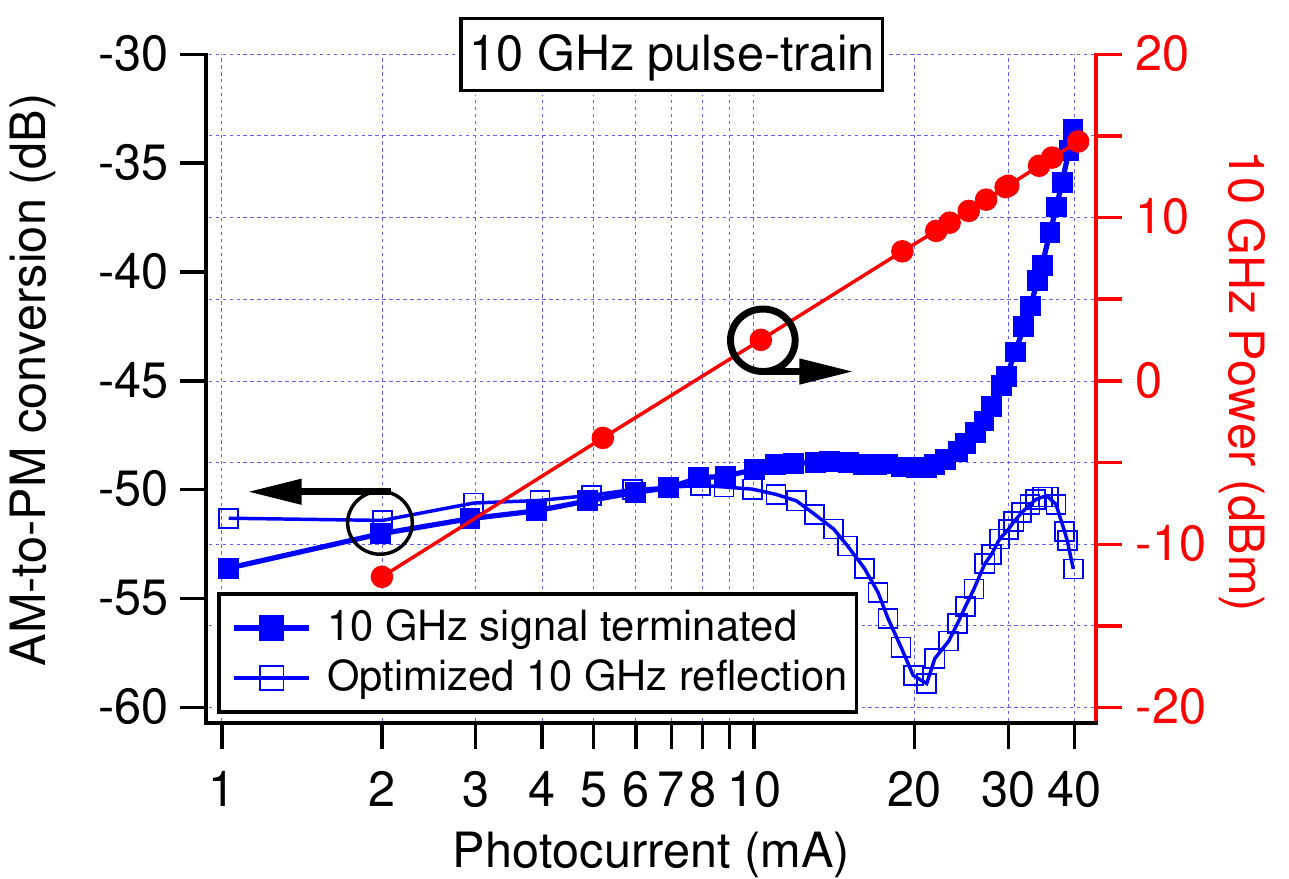}
\caption{\mbox{AM-to-PM} coefficient and microwave power using a 10~GHz pulse-train. When a small amount of 10~GHz reflection is allowed, we can reduce the \mbox{AM-to-PM} coefficient and maintain it below -50~dB out to 40~mA of photocurrent (blue empty squares). The power measurement shown in red (right axis) is calibrated to the connector at the bias tee.}
\label{fig:results3}
\end{figure}

To test the photodiode under the most favorable conditions, we use a second optical pulse source consisting of a 10~GHz frequency comb generated via electro-optic modulation of continuous-wave light \cite{Beha2017,Metcalf2013}. The spectral width of this comb is $\sim$1 THz and the pulses are also compressed to nearly transform limited pulse-width of $< 2$~ps before being sent to the photodiode. Using a pulse-train directly at 
10~GHz has the advantage of having a reduced energy-per-pulse, which extends the linear regime of the photodiode to higher photocurrents \cite{Hu2017}. When using the 10~GHz electro-optic frequency comb to measure the \mbox{CC-MUTC's} \mbox{AM-to-PM} coefficient, we find that the 20~GHz harmonic (uncontrolled in our current setup) plays a more prominent role. The importance of the uncontrolled 20~GHz harmonic rejection is indicated by the shape of the \mbox{AM-to-PM} curve in solid squares in Fig. 5 where the curve does not present a clear null. Similar to Fig. 3, the lack of a null in the \mbox{AM-to-PM} curve signifies the presence of nonlinear distortion due to microwave reflection. By controlling the reflection of a small fraction of the 10~GHz output, we obtain \mbox{AM-to-PM} coefficients below -50~dB out to 40~mA, as shown in Fig. 5. At such photocurrent, the microwave power in the 10~GHz harmonic at the bias tee is $\sim$15~dBm and we estimate that the microwave power at the photodiode surpasses 17~dBm. The possibility of producing such high microwave powers while operating with sufficient linearity enables interesting possibilities in microwave beam forming with fiber-fed remote antennas \cite{Frankel1997}.

\section{Comparisons and Discussion}
To place the extremely low nonlinearity level of \mbox{CC-MUTC} photodiodes in context, we have compared their performance to that of p-i-n photodiodes with both standard and high-linearity designs. Two representative cases \cite{Taylor2011,Joshi2008} are shown in Fig. 6. We emphasize the fact that not only are \mbox{CC-MUTC} photodiodes capable of operating with extremely low \mbox{AM-to-PM} coefficients, but they also do so over an extended photocurrent range, allowing their application in varied conditions without the need to fine tune or stabilize the optical power. Furthermore, the \mbox{AM-to-PM} can stay at low values out to $> 40~\textrm{mA}$ of photocurrent, an unprecedented number even compared to other MUTC designs.

\begin{figure}[htbp]
\centering
\includegraphics[width=3.3 in]{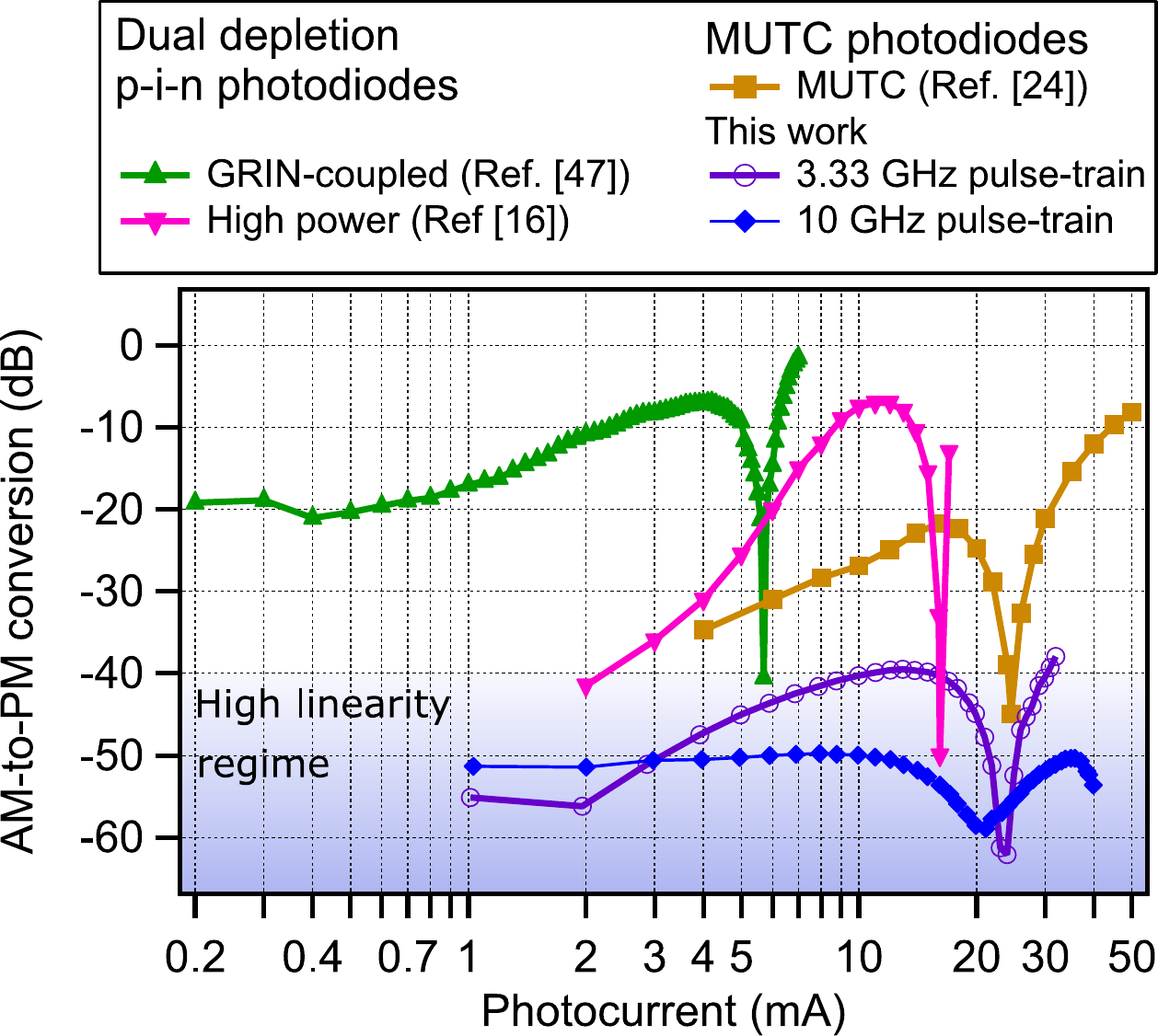}
\caption{Nonlinear performance comparison between leading photodiode structures. The blue range shows values of \mbox{AM-to-PM} typically considered to be acceptable. Notice how in standard photodiodes this range is only reached in a narrow photocurrent range.}
\label{fig:comparison}
\end{figure}

One of the applications with the most stringent requirements on the photodetector linearity under ultrashort pulse illumination is the generation of low noise microwave signals using OFD. To assess the performance of these diodes, we must determine a reasonable requirement for the \mbox{AM-to-PM} coefficient. Taking as typical an Er:fiber based mode-locked laser as a frequency divider with a RIN plateau at $\sim$-140~dBc/Hz \cite{Chen2007}, a RIN-to-phase noise rejection of 40~dB is sufficient to support the division of state-of-the-art cavity stabilized lasers. With a RIN rejection of $> 50$~dB as demonstrated here, the \mbox{AM-to-PM} limits the phase noise performance only at -190~dBc/Hz, a level not yet demonstrated in any photodetected signal, thereby supporting next-generation experiments in microwave generation via OFD. After suppression of RIN-induced phase noise, other sources such as shot noise \cite{Quinlan2013} and photocarrier scattering in the photodiode \cite{Sun2014} limit the overall performance.

In the more general case of microwave photonic links, the nonlinear impairments inherent in the optical-to-electrical conversion can be improved with the methods demonstrated here. We expect that microwave reflections to the photodiode will impact all IMD products in microwave photonic links and not only when all signals are harmonically related. For example, in many microwave photonic links it is the third-order nonlinearity that is most important, since the second-order sum or difference frequencies are generally out of band. However, controlled reflection of the second-order nonlinear products may help suppress third-order nonlinearities, improving link performance. Although we have focused on reducing the nonlinearity, reflecting back microwave signals can also increase the nonlinear response. This can be useful in optoelectronic mixing applications where a large photodetector nonlinearity is desired to perform up- or down-conversion of photonically encoded microwave signals \cite{Pan2010}. Finally, in OEOs, the \mbox{AM-to-PM} coefficient has been shown to degrade the OEO performance \cite{Eliyahu2008}. We expect that the high-linearity photodiodes and broadband impedance control demonstrated here will also impact the noise performance of OEOs.

In conclusion, we have shown that high linearity \mbox{CC-MUTC} photodiodes in conjunction with broadband impedance control of the external circuit can perform optical-to-electrical conversion with better than 50~dB rejection of \mbox{AM-to-PM} conversion over a very large range of photocurrents. When illuminated at a pulse repetition rate of 10~GHz, 
15~dBm of microwave power is generated with an \mbox{AM-to-PM} coefficient of -53~dB. The generation of still higher power microwaves with high linearity appears feasible. This performance level is likely to impact photonic generation of low noise microwaves as well as microwave photonic links.

\textbf{Funding.} Defense Advanced Projects Agency (DARPA) (PULSE program); Naval Research Laboratory (NRL); National Institute of Standards and Technology (NIST)

\textbf{Acknowledgments.} We thank N. Orloff and J. Booth for assistance in the early stage of this work, D. Carlson for the use of the 10~GHz comb source, and D. Nicolodi and A. Kowligy for useful comments on the manuscript. This work is a contribution of an agency of the US government and not subject to copyright in the USA.

\bibliographystyle{IEEE}
\bibliography{AMPMPaperBibliography}

\appendix
\counterwithin{figure}{section}
\section{Supplementary Information}
The goal of this supplement is to develop a simple model which enables quantitative predictions of the nonlinearity modification in a fast photodiode when part of the generated microwave signal is reflected back to the diode. This model predicts the trends seen in Figs. 3 to 5 of the main text, as well as predicting only small shifts in the \mbox{AM-to-PM} null for photodiodes with high nonlinearity.

In photodiode intermodulation distortion, microwave tones generated via illumination of a modulated optical signal nonlinearly mix, producing sum and difference frequencies, third order intermodulation products, and higher-order terms\cite{Williams1996}. Importantly, in the photodetection of ultra-short pulses, all the microwave tones present at the photodiode are harmonics of the pulse-train repetition frequency, permitting the existence of several combinations of harmonics which yield a term at the microwave frequency of interest, as illustrated in Fig. 1(e) in the main text. For example, consider an optical pulse-train with repetition frequency $f$, generating microwave signals at $f$, $2f$, $3f$, ..., $Nf$ at the photodiode. If the microwave frequency of interest is $3f$, the photodiode’s second-order nonlinearity can generate the sum frequency of $f$ and $2f$, which adds to the microwave signal at frequency $3f$. These terms will interfere, modifying the amplitude and phase of the microwave signal at $3f$. A similar effect takes place when a microwave signal is introduced along the electrical path, modulating the diode's bias voltage\cite{Cabon2010,Jaro2003}. The diode responds to the bias modulation according to the exponential current-voltage relation. A Taylor series expansion of the current-voltage relation around the static bias voltage reveals terms to all orders in the perturbation voltage.

Under the assumption that the perturbation remains small, only the first few terms need to be considered. Thus, the introduction of one or more microwave signals through the voltage bias results in the generation of sum and difference frequencies, third order intermodulation products and higher-order terms, that interfere with the intermodulation products directly generated in the photodiode. As shown in the results in the main text, this newly generated term can significantly alter the nonlinear behavior of the diode. 

To further elucidate this concept, consider a new tone at the microwave frequency of interest generated via nonlinear mixing of a reflected tone, added to the original microwave signal:
\begin{equation}
s(t) = \sin(x) + A \sin(x + \phi),
\end{equation}
where $A$ and $\phi$ represent the relative amplitude and phase between the nonlinearly generated harmonic and the directly generated microwave. The sum of these voltages can be rewritten as
\begin{equation}
s(t) = \sqrt{1 + 2A\cos\phi +A^2}\sin(x + \theta),
\end{equation}
resulting in a signal phase shifted by an angle $\theta$, which depends both on $A$ and $\phi$,
\begin{equation}
\theta = \tan^{-1}\left[\frac{A\sin\phi}{1 + A\cos\phi}\right].
\end{equation}

Since the phase of the final tone at the frequency of interest has an amplitude dependence, any nonlinear response of A as a function of photocurrent will generate an \mbox{AM-to-PM} term in addition to the photodiode’s native nonlinearity. To evaluate the \mbox{AM-to-PM} contribution from this term, we need to evaluate how it changes with photocurrent, that is,
\begin{equation}
\frac{\textrm{d}\theta}{\textrm{d}I} = \frac{\textrm{d}\theta}{\textrm{d}A}\cdot\frac{\textrm{d}A}{\textrm{d}I}.
\end{equation}

Evidently, if the relative amplitude of the additional microwave term has any dependence on the photocurrent (that is, if $\textrm{d}A/\textrm{d}I\neq 0$), there will be an additional amplitude-to-phase conversion term. To evaluate how this function affects the overall nonlinearity, we evaluate the simplest model where the power of the generated microwaves is quadratic with photocurrent, making the power in the tone generated via 2nd order nonlinear mixing dependent on the fourth power of the photocurrent. Therefore, the relative amplitude of the newly generated to the original microwave is proportional to the photocurrent,
\begin{equation}
A(I) = \beta I,
\end{equation}
where $\beta$ is a constant. Physically, $\beta^{-1}$ is the (extrapolated) photocurrent at which the tone generated by nonlinear mixing has the same amplitude as the photo-generated microwave signal. Most likely our model breaks down before we reach such a regime, since at this point higher order nonlinear terms should also be included.

Using Eqs. (S3) to (S5), we can calculate the additional \mbox{AM-to-PM} by obtaining the derivative of the overall phase as a function of photocurrent,
\begin{equation}
\frac{\textrm{d}\theta}{\textrm{d}I/I} = 
\frac{A\sin\phi}{1+2A\cos\phi+A^2}.
\end{equation}

A few interesting features are: 1) this function is periodic with $\phi$, the phase of the newly generated wave, a parameter controlled by the delay of the reflection, 2) the effect has zero amplitude for $\phi = n\pi$ and, 3) the sign of this added \mbox{AM-to-PM} term can change with the phase $\phi$, permitting control of how it affects the diode's inherent nonlinearity.

\begin{figure}[htbp]
\centering
\includegraphics[width=3.3 in]{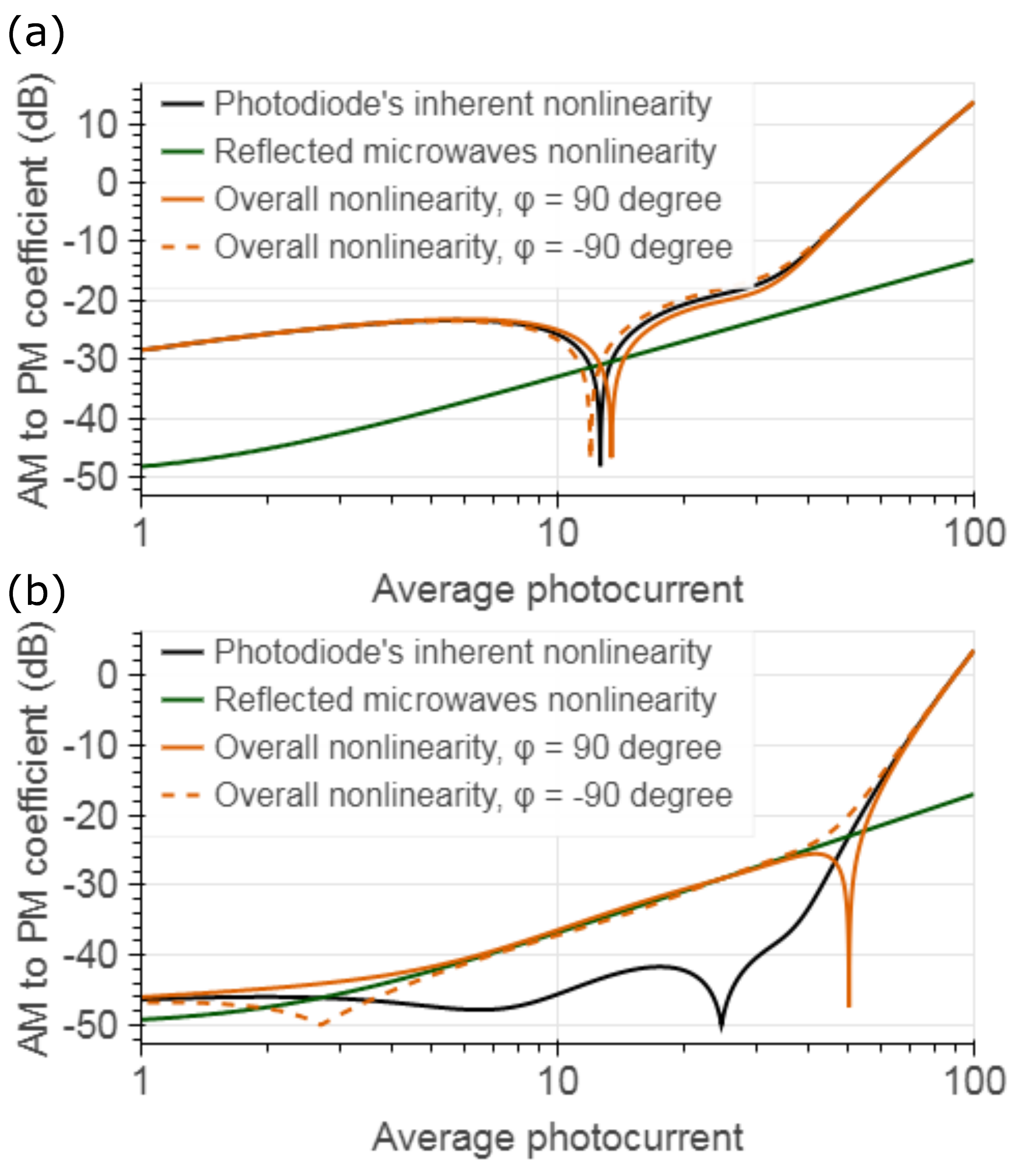}
\caption{A simulated \mbox{AM-to-PM} coefficient and its modification by microwave mixing at the diode. The purple solid and dashed curves show the cases when the nonlinearity adds with opposite signs respectively. (a) shows ``typical" photodiode \mbox{AM-to-PM}. Notice that in (a), the modification only induces a slight variation in the location of the \mbox{AM-to-PM} null. (b) shows the \mbox{AM-to-PM} in a high linearity photodiode and its modification.}
\label{fig:suppl1}
\end{figure}

To illustrate how this nonlinearity affects the performance of a photodiode system, we've chosen two cases, shown in Fig. A1. Fig. A1 (a) shows the case of a ``typical" photodiode with a narrow null and an otherwise high \mbox{AM-to-PM} coefficient. For maximum effect, we have chosen the phase of the reflected signal to be $\pm 90$ degrees. As is evident, this additional nonlinearity only changes the \mbox{AM-to-PM} value near the null. On the other hand, for the case of a photodiode with high linearity, as shown in Fig. A1 (b), the additional nonlinearity due to microwave reflections can have significant impact and it can dominate the overall \mbox{AM-to-PM}. In addition, this nonlinearity modification can create a null at a high photocurrent due to a cancellation between the diode’s inherent nonlinearity and the additional nonlinearity described here.

To compare with the experimental results presented in Fig. 3 in the main text, we can also use this model to generate the modified nonlinearity for the case of a highly linear photodiode and for a few values of the phase of the reflected signal. The simulated \mbox{AM-to-PM} is shown in Fig. A2. As the value of the reflected phase increases from 0, the null generated due to the cancellation moves to higher photocurrents, as observed in the experiment. When the phase is $<0$ then the modified \mbox{AM-to-PM} curve does not exhibit a null as the sign of the nonlinearity does not permit a cancellation. This effect is also seen in Fig. 3 of the main text.

\begin{figure}[htbp]
\centering
\includegraphics[width=3.3 in]{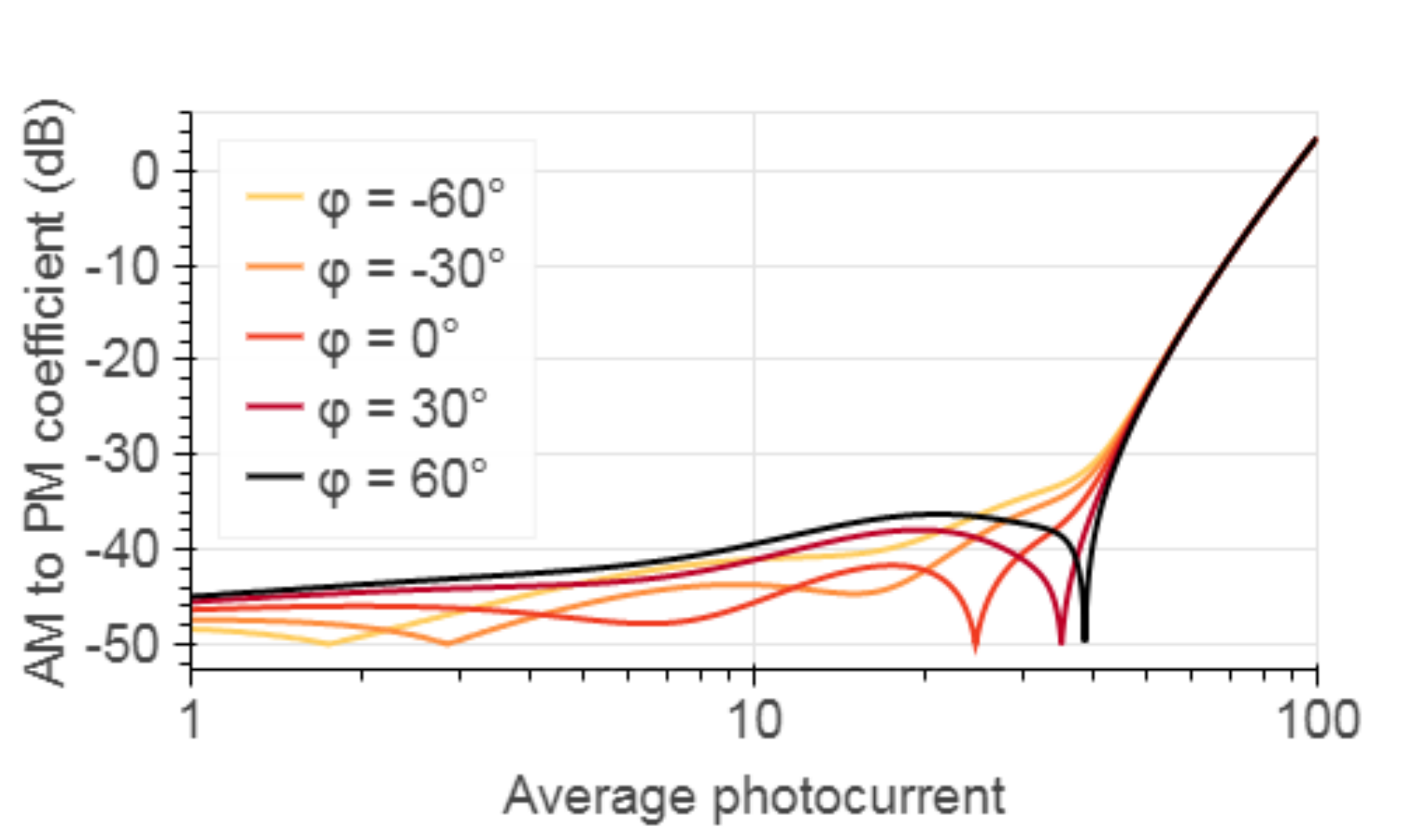}
\caption{Modification of the \mbox{AM-to-PM} coefficient for a few chosen values of the reflected phase.}
\label{fig:suppl2}
\end{figure}

\end{document}